\documentclass[preprint,showpacs,preprintnumbers,amsmath,amssymb]{revtex4}

\usepackage{graphicx}
\usepackage{dcolumn}
\usepackage{bm}
\usepackage{amssymb}

\begin{document}
\title{Classical small systems coupled to finite baths
}

\author{Hideo Hasegawa}
\altaffiliation{hideohasegawa@goo.jp}
\affiliation{Department of Physics, Tokyo Gakugei University,  
Koganei, Tokyo 184-8501, Japan}%

\date{\today}

\begin{abstract}
We have studied the properties of a classical $N_S$-body system
coupled to a bath containing
$N_B$-body harmonic oscillators, employing an $(N_S+N_B)$ model which is
different from most of the existing models with $N_S=1$.
We have performed simulations for $N_S$-oscillator systems,
solving $2(N_S+N_B)$ first-order differential equations with $N_S \simeq 1 - 10$ and
$N_B \simeq 10 - 1000$, in order to calculate the time-dependent 
energy exchange between the system and the bath.
The calculated energy in the system rapidly changes
while its envelope has a much slower time dependence.
Detailed calculations of the stationary energy distribution of the system $f_S(u)$
($u$: an energy per particle in the system) have shown that its properties are
mainly determined by $N_S$ but weakly depend on $N_B$.
The calculated $f_S(u)$ is analyzed with the use of
the $\Gamma$ and $q$-$\Gamma$ distributions:  
the latter is derived with the superstatistical approach (SSA)
and microcanonical approach (MCA) to the nonextensive statistics,
where $q$ stands for the entropic index.
Based on analyses of our simulation results,
a critical comparison is made between the SSA and MCA.
Simulations have been performed also for the $N_S$-body ideal-gas system.
The effect of the coupling between oscillators in the bath has been examined
by additional ($N_S+N_B$) models which include baths consisting of 
coupled linear chains with periodic and fixed-end boundary conditions.

\end{abstract}

        

\maketitle
\newpage
\section{Introduction}

The study on open systems is one of the important areas in
classical and quantum statistics \cite{Weiss99}.
In the theory of open systems, the deterministic dynamics of
particles in the system is replaced by the
stochastic Langevin equation in the classical limit.
The problem has been investigated with the use of 
various models in which a single particle (the system)
is attached at the center (or edge) of a linear chain \cite{Ford65,Plyukhin01}, or
it is coupled to a bath consisting of a collection
of harmonic oscillators 
\cite{Ullersma66}-\cite{Rosa08}.
Many studies have been made for open systems by using
the Caldeira-Leggett (CL) model given by \cite{Ullersma66,Caldeira81,Caldeira83}
\begin{eqnarray}
H_{CL} = \frac{P^2}{2M}+ V(Q)
+ \sum_{n=1}^{N_B} \left[ \frac{p_n^2}{2m}+
\frac{m \omega_n^2}{2}\left(q_n - \frac{c_n}{m\omega_n^2}Q \right)^2
\right],
\label{eq:I1}
\end{eqnarray}
where $M$ ($m$), $P$ ($p_n$) and $Q$ ($q_n$) denote the mass, 
momentum and coordinate of a particle in a system (bath), 
$V(Q)$ stands for the potential in the system,
$\omega_n$ the frequency of the $n$th oscillator in the $N_B$-body bath 
and $c_n$ the coupling constant between the system and bath.
The CL model was originally introduced for infinite bath
($N_B \rightarrow \infty$). In recent years,
the CL model has been employed for a study of properties of a small system 
coupled to a {\it finite} bath \cite{Smith08}-\cite{Rosa08}.
A thermalization of a particle (the system) coupled to a finite bath 
has been investigated \cite{Smith08,Wei09}.
It has been shown that a complete thermalization of the
particle requires some conditions for relative ranges of
oscillating frequencies in the system and bath
\cite{Smith08,Wei09}.
The specific heat of a single oscillator (the system) coupled to finite bath
has been studied with the use of two different evaluation methods
\cite{Hanggi08,Ingold09}.
The energy exchange between particles in a rachet potential
(the system) and finite bath ($N_B=1-500$) has been investigated
\cite{Rosa08}.

Ford and Kac proposed the model given by \cite{Ford87}
\begin{eqnarray}
H_{FK} = \frac{P^2}{2M}+ V(Q)
+ \sum_{n=1}^{N_B} \left[ \frac{p_n^2}{2m}+
\frac{m \omega_n^2}{2}\left(q_n -Q \right)^2
\right],
\label{eq:J1}
\end{eqnarray}
which is referred to as the FK model.
The CL and FK models are formally equivalent \cite{Ford87}
because Eq. (\ref{eq:J1}) may be derived from Eq. (\ref{eq:I1}) 
with $c_n = m \omega_n^2$. 
However, the physical meanings of the coupling term in the CL
and MK models are not the same. 
The CL model was initially introduced such that we take into account 
a linear coupling of $- Q \sum_n c_n q_n$ between system and bath \cite{Ullersma66},
and then the counter term of $c_n^2 Q^2/m \omega_n^2$ was included 
for a compensation of the renormalization in
the oscillating frequency by the introduced interaction.
In contrast, the interaction term in Eq. (\ref{eq:J1}) of the FK model 
clearly expresses the quadratic potential of springs between $Q$ and $q_n$.
It is evident that the interaction term of the FK model in Eq. (\ref{eq:J1})
preserves the translational invariance whereas that of the CL model 
in Eq. (\ref{eq:I1}) does not in a strict sense 
\cite{Hakim85,Canizares94,Patriarca96} except for $c_n = m \omega_n^2$ 
for which the CL model reduces to the FK model as mentioned above. 
The importance of the translationally invariant interaction 
in the system plus bath models has been discussed 
in Refs. \cite{Hakim85,Canizares94,Patriarca96}.

In existing models which have been proposed for open systems 
\cite{Caldeira81}-\cite{Rosa08},
the number of particles in a systems is taken to be unity ($N_S=1$) while
a generic open system may contain any number of particles.
It is necessary 
to develop an ($N_S+N_B$) model including {\it finite} $N_S$-body system ($N_S \geq 1$)
coupled to $N_B$-body bath, 
with which we may investigate the properties of generic small systems.
Extending the FK model, we will propose in this paper
three types of ($N_S+N_B$) models (referred to as A, B and C).
In the model A a bath consists of uncoupled oscillators, and
in the models B and C baths contain coupled oscillators
with the periodic and fixed-end boundary conditions, respectively.
They are adopted for a study on effects of couplings in bath oscillators.

In the last decade, many studies have been made for nonextensive
statistics initially proposed by Tsallis \cite{Tsallis88}-\cite{Tsallis04}.
In nonextensive systems, the probability distribution generally does not follow
the Gaussian, but it is well described by the $q$-exponential distribution,
\begin{eqnarray}
p(u) &\propto& e_q^{-\beta_0 u} = [1-(1-q)\beta_0 u]_{+}^{1/(1-q)},
\label{eq:I2}
\end{eqnarray}
where an inverse of the effective temperature $\beta_0$ and 
the entropic index $q$ are fitting parameters, and
the $q$-exponential function $e_q^x$ is defined by 
\cite{Tsallis88}-\cite{Tsallis04}
\begin{eqnarray}
e_q^{x} &=& [1+(1-q)x]_{+}^{1/(1-q)}, 
\label{eq:L5}
\end{eqnarray}
with $[y]_{+}={\rm max}(y, 0)$. In the limit of $q \rightarrow 1.0$, 
$e_q^{x}$ reduces to the exponential function $e^x$.
In a seminal paper \cite{Tsallis88}, 
the $q$-exponential distribution was first derived by the maximum-entropy method
with the use of the so-called Tsallis entropy.
Later superstatistical \cite{Wilk00,Beck01} and microcanonical methods 
\cite{Plastino94,Almeida01} have been proposed
as alternative approaches to nonextensive statistics.
Recent development has shown that small systems belong to nonextensive
systems \cite{Tsallis04}.
Performing direct simulation (DS) for the proposed ($N_S+N_B$) model
with the system containing independent $N_S$ oscillators (the model A), 
we have calculated the stationary distribution of the system of $f_S(u)$ 
for the energy per particle $u$ ($=E_S/N_S$, $E_S$: the system energy). 
The calculated distribution is well described by the $q$-$\Gamma$ distribution
given by
\begin{eqnarray}
f_S(u) &\propto& u^{a-1} e_q^{-b u},
\label{eq:L9}
\end{eqnarray}
where $a$, $b$ and $q$ are fitting parameters.
It is easy to see that in the limit of $q \rightarrow 1.0$, 
the $q$-$\Gamma$ distribution 
reduces to the conventional $\Gamma$ distribution.
As will be shown in Sec. III,
superstatistical approach (SSA) \cite{Wilk00,Beck01}
and microcanonical approach (MCA) \cite{Plastino94}-\cite{Aringazin03} 
lead to the equivalent expressions for $f_S(u)$ 
given by Eq. (\ref{eq:L9}) with $a=N_S$ and $b= \beta_0 N_S$,
but with different expressions
for the entropic index $q$: 
\begin{eqnarray}
q &= & \left\{ \begin{array}{ll}
1+\frac{1}{N_S} \geq 1.0
\quad & \mbox{in the SSA}, \\
1-\frac{1}{(N_B-1)} \leq 1.0
\quad & \mbox{in the MCA}. 
\end{array} \right.
\label{eq:L10}
\end{eqnarray}
The entropic index in the SSA is expressed in terms of a system parameter ($N_S$),
while that in the MCA is expressed in terms of a bath parameter ($N_B$).
This difference is serious from the physical viewpoint of small open systems.
The purpose of the present paper is twofold: to develop the ($N_S+N_B$) model
in which an open system contains finite $N_S$ particles, and to investigate
the validity of the stationary distribution functions derived in the SSA 
\cite{Wilk00,Beck01} and MCA \cite{Plastino94}-\cite{Aringazin03}.
This is the first study on open systems with finite $N_S$ ($\geq 1$)
as far as we are aware of.

The paper is organized as follows.
In Sec. II, we propose the model A mentioned above, for which
we perform DS of $2(N_S+N_B)$ differential equations
for the $N_S$-oscillator system in order to calculate the time-dependent 
energy exchange between the system and bath.
We present detailed calculations of $f_S(u)$, 
changing model parameters such as $N_S$, $N_B$, frequency distribution,
mass of oscillators in the bath, 
and coupling strength between the system and bath.
In Sec. III, we analyze the calculated $f_S(u)$ 
by using the $\Gamma$ distribution [Eq. (\ref{eq:D2})]
and the $q$-$\Gamma$ distribution [Eq. (\ref{eq:L9}) or (\ref{eq:L4})].
The former is derived based on the Boltzmann-Gibbs statistics 
and the latter is obtained with the SSA \cite{Wilk00,Beck01}
and MCA \cite{Plastino94}-\cite{Aringazin03} of the nonextensive statistics.
DS has been made also for the system consisting of $N_S$-body ideal gases,
whose results are compared to those of oscillators.
We introduce the models B and C, whose DS for the oscillator systems
will be reported.
A comparison is made among Langevin equations derived 
in various models for open systems.
The final Sec. IV is devoted to our conclusion.

\section{Adopted ($N_S+N_B$) model}

\subsection{A system with bath containing uncoupled oscillators}
We consider a system ($H_S$) and a bath ($H_B$) consisting of 
independent $N_S$ and $N_B$ one-dimensional oscillators, respectively,
which are coupled by the interaction ($H_{I}$).
We assume that the total Hamiltonian is given by
\begin{eqnarray}
H &=& H_S+H_B+H_{I},
\label{eq:A0}
\end{eqnarray}
with
\begin{eqnarray}
H_S &=& \sum_{k=1}^{N_S} \left[ \frac{P_k^2}{2M}+ V(Q_k) \right]
- f(t) \sum_{k=1}^{N_S} Q_k, 
\label{eq:A2}\\
H_B &=& \sum_{n=1}^{N_B} \left[ \frac{p_n^2}{2m}+ v(q_n) \right], 
\label{eq:A3}\\
H_{I} &=& \frac{1}{2} \sum_{k=1}^{N_S}  \sum_{n=1}^{N_B} 
c_{kn} (Q_k-q_n)^2, 
\label{eq:A4}\\
v(q_n) &=& \frac{b_n}{2} q_n^2=\frac{m \omega_n^2}{2} q_n^2
\hspace{2cm}\mbox{(model A)},
\label{eq:A1}
\end{eqnarray}
which is referred to as the model A.
Here $M$ ($m$) denotes the mass, $P_k$ ($p_n$) the momentum, 
$Q_k$ ($q_n$) position of the oscillator,
$V(Q_k)$ ($v(q_n)$) the potential in the system (bath),
$c_{nk}$ coupling constant, $b_n$ and $\omega_n$ 
spring constant and frequency in the bath, respectively, 
and $f(t)$ an external force.
A simple generalization of the FK model [Eq. (\ref{eq:J1})]
yields the model Hamiltonian given by Eq. (\ref{eq:A0}) with 
$H_S$ given by Eq. (\ref{eq:A2}),
$H_B =\sum_{n=1}^{N_B} p_n^2/2 m$ and 
$H_I = \sum_{k=1}^{N_S} \sum_{n=1}^{N_B} (m \omega^2/2)(q_n-Q_k)^2$.
In our model Hamiltonian,
we have added $v(q_n)$ in $H_B$ such that the Hamiltonian
is symmetric with respect
to an exchange of system $\leftrightarrow$ bath (for $f(t)=0$)
and such that we may discuss the coupled oscillators in baths (model B and C). 
Furthermore, we have included coupling $c_{kn}$ in place of $m \omega^2$ in $H_I$
of the generalized FK model in order to study the effect of system-bath couplings.
We note that $H_I$ in Eq. (\ref{eq:A4}) may be rewritten as
\begin{eqnarray}
H_I &=& \frac{1}{2} \sum_{k=1}^{N_S} \left(\sum_{n=1}^{N_B} c_{kn} \right) Q_k^2
+ \frac{1}{2} \sum_{n=1}^{N_B} \left(\sum_{k=1}^{N_S} c_{kn} \right) q_n^2
-  \sum_{k=1}^{N_S} \sum_{n=1}^{N_B} c_{kn} Q_k q_n.
\label{eq:A30}
\end{eqnarray}
Absorbing the first and second terms in Eq. (\ref{eq:A30}) to $H_S$ and  $H_B$,
respectively, we may regard the last term as the interaction.
Such a model Hamiltonian with a linear coupling of $-\sum_k \sum_n c_{kn} Q_k q_n$
corresponds to the generalized CL model for finite $N_S$.

From Eqs. (\ref{eq:A0})-(\ref{eq:A1}), we obtain $2(N_S+N_B)$ first-order
differential equations, 
\begin{eqnarray}
\dot{Q}_k &=& \frac{P_k}{M}, 
\label{eq:A10b}\\
\dot{P}_k 
&=& - V'(Q_k)-\sum_{n=1}^{N_B} c_{kn} (Q_k-q_n) +f(t), \\
\dot{q}_n &=& \frac{p_n}{m}, \\
\dot{p}_n
&=& -m \omega_n^2 q_n- \sum_{k=1}^{N_S} c_{kn} (q_n-Q_k),
\label{eq:A11b}
\end{eqnarray}
which yield
\begin{eqnarray}
M \ddot{Q}_k 
&=& - V'(Q_k)-\sum_{n=1}^{N_B} c_{kn} (Q_k-q_n) +f(t), 
\label{eq:A10}\\
m \ddot{q}_n
&=& -m \omega_n^2 q_n- \sum_{k=1}^{N_S} c_{kn} (q_n-Q_k),
\label{eq:A11}
\end{eqnarray}
prime ($'$) and dot ($\cdot$) denoting derivatives with respect 
to the argument and time, respectively.

A formal solution of Eq. (\ref{eq:A11}) for $q_n(t)$ is given by
\begin{eqnarray}
q_n(t) &=& q_n(0) \cos \tilde{\omega}_n t 
+ \frac{\dot{q}_n(0)}{\tilde{\omega}_n} \sin \tilde{\omega}_n t 
+ \sum_{\ell=1}^{N_S} \frac{c_{\ell n}}{m \tilde{\omega}_n} \int_0^t 
\sin \tilde{\omega}_n (t-t') Q_{\ell}(t')\:dt',
\label{eq:A12}
\end{eqnarray}
with
\begin{eqnarray}
\tilde{\omega}_n^2 &=&  \frac{b_n}{m}+\sum_{k=1}^{N_S} \frac{c_{kn}}{m}
=\omega_n^2 +\sum_{k=1}^{N_S} \frac{c_{kn}}{m}.
\end{eqnarray}
Substituting Eq. (\ref{eq:A12}) to Eq. (\ref{eq:A10}), 
we obtain the Langevin equation given by 
\begin{eqnarray}
M \ddot{Q}_k(t) &=& -V'(Q_k)- M \sum_{\ell=1}^{N_S} \xi_{k\ell} Q_{\ell}(t)
- \sum_{\ell=1}^{N_S} \int_0^t \gamma_{k\ell}(t-t') \dot{Q}_{\ell}(t')\:dt' 
\nonumber \\
&-& \sum_{\ell=1}^{N_S} \gamma_{k\ell}(t) Q_{\ell}(0) + \zeta_k(t)+f(t)
\hspace{1cm}\mbox{($k =1$ to $N_S$)},
\label{eq:A13}
\end{eqnarray} 
with
\begin{eqnarray}
M \xi_{k\ell} &=& \sum_{n=1}^{N_B}  \left[ c_{kn}  \delta_{k \ell} 
-\frac{c_{kn} c_{\ell n}}{m \tilde{\omega}_n^2} \right], 
\label{eq:A14}\\
\gamma_{k\ell}(t) 
&=&\sum_{n=1}^{N_B} \left( \frac{c_{kn} c_{\ell n}}{m \tilde{\omega}_n^2}\right) 
\cos \tilde{\omega}_n t, 
\label{eq:A15}\\
\zeta_k(t) &=& \sum_{n=1}^{N_B} c_{kn} 
\left[q_n(0) \cos \tilde{\omega}_n t
+ \frac{\dot{q}_n(o)}{\tilde{\omega}_n} \sin \tilde{\omega}_n t \right],
\label{eq:A16}
\end{eqnarray}
where $\xi_{k \ell}$ denotes the additional interaction between
$k$ and $\ell$th particles in the system induced by couplings $\{ c_{kn} \}$, 
$\gamma_{k\ell}(t)$ the memory kernel and $\zeta_k$ the stochastic force.

If the equipartition relation is realized 
in initial values of $q_n(0)$ and $\dot{q}(0)$,
\begin{eqnarray}
\langle m \tilde{\omega}_n^2 q_n(0)^2\rangle_B
&=& \langle m  \dot{q}_n(0)^2\rangle_B = k_B T,
\label{eq:A17}
\end{eqnarray} 
we obtain the fluctuation-dissipation relation:
\begin{eqnarray}
\langle \zeta_k(t) \zeta_k(t') \rangle_B &=& k_B T \gamma_{kk}(t-t'),
\label{eq:A18}
\end{eqnarray}
where $\langle \cdot \rangle_B$ stands for the average
over variables in the bath.

In the case of $N_B \rightarrow \infty$, summations 
in Eqs. (\ref{eq:A14})-(\ref{eq:A16}) are replaced by integrals. 
When the density of states 
($D(\omega)= N_B^{-1} \sum_n \delta(\omega-\omega_n)$)
is given by the Debye form: $D(\omega) \propto \omega^2$ for $0 \leq \omega < w_D$, 
the kernel becomes
\begin{eqnarray}
\gamma(t) \propto \frac{\sin \omega_D t}{\pi t} \propto \delta(t),
\end{eqnarray}
which leads to the Markovian Langevin equation.

In the case of $N_S=1$, we obtain $\xi$ and $\gamma$ 
in Eqs. (\ref{eq:A14}) and (\ref{eq:A15}) 
where the subscripts $k$ and $\ell$ are dropped ({\it e.g.,} $c_{kn}=c_n$),
\begin{eqnarray}
M \xi(t) &=& \sum_{n=1}^{N_B} c_n \left( 1-\frac{c_n}{m \tilde{\omega}_n^2} \right), 
\label{eq:A19}\\
\gamma(t) &=& \sum_{n=1}^{N_B} \left( \frac{c_n^2}{m \tilde{\omega}_n^2}\right) \cos \tilde{\omega}_n t.
\label{eq:A20}
\end{eqnarray}
The additional interaction vanishes ($\xi=0$)
if we choose $c_n=m \tilde{\omega}_n^2$ in Eq. (\ref{eq:A19}). 

In the case of $N_S \neq 1$, however, it is impossible to choose
$\{ c_{kn} \}$ such that $\xi_{k \ell}=0$ is realized for all pairs of $(k, \ell)$
in Eq. (\ref{eq:A14}).
Then $Q_k$ is inevitably coupled to $Q_{\ell}$ for $\ell \neq k$ 
with the superexchange-type interaction of antiferromagnets: 
$-\sum_n c_{kn}c_{\ell n}/m \tilde{\omega}_n^2$ in Eq. (\ref{eq:A14}).

\subsection{Model calculations for oscillator systems}

It is easier to solve $2(N_S+N_B)$ first-order differential equations
given by Eqs. (\ref{eq:A10b})-(\ref{eq:A11b}) 
than to solve the $N_S$ Langevin equations given by Eqs. (\ref{eq:A13})-(\ref{eq:A16})
although the latter provides us with clearer physical insight than the former. 
We have performed DS, solving the differential equations
for the oscillator system with $V(Q_k)=M \Omega_k^2 Q_k^2/2$ in Eq. (\ref{eq:A2})
for $f(t)=0$, $M=m=1.0$ and $\Omega_k =\omega_n=1.0$ otherwise noticed 
with the use of the fourth-order Runge-Kutta method with the time step of 0.01.
In order to study the $N_S$ and $N_B$ dependences of various physical quantities,
we have assumed the coupling given by 
\begin{eqnarray}
c_{k n}=\frac{c_0}{N_S N_B},
\label{eq:E1}
\end{eqnarray}
because the interaction term includes summations of 
$\sum_{k=1}^{N_S}$ and $\sum_{n=1}^{N_B}$ in Eq. (\ref{eq:A4}).
We have chosen $c_0=10.0$ (see Sec. II B 3. {\it Effect of} $c_0$). 
It is noted that with our choice of $c_{k n}$, the interaction contribution
is finite even in the thermodynamical limit of $N_B \rightarrow \infty$ because
the summation over $n$ runs from 1 to $N_B$ in Eq. (\ref{eq:A4}).
Although we have tried to adopt an alternative choice of $c_{k n}$ 
given by \cite{Note2}
\begin{eqnarray}
c_{k n}=\frac{c_0'}{\sqrt{N_S N_B} },
\label{eq:E2}
\end{eqnarray}
qualitatively similar results have been obtained, as will be shown in Sec. III A.
Initial conditions for $Q_k(0)$, $\dot{Q}_k(0)$, $q_n(0)$ and $\dot{q}_n(0)$
are given by random Gaussian variables with zero means and unit variances.
Simulations have been performed for $t=0$ to 10000,
results for $t < 2000$ being discarded for evaluations of stationary distributions.
Results to be reported are averages over 10000 runs.

We have assumed that the energies per particle $u_{\eta}(t)$ 
in the system ($\eta$=S) and the bath ($\eta$=B) are given by
\begin{eqnarray}
u_S &=& \frac{1}{N_S} \sum_{k=1}^{N_S} \left[\frac{P_k^2}{2 M}
+ \frac{M \Omega_k^2 Q_k^2}{2} \right], \\
u_B &=& \frac{1}{N_B} \sum_{n=1}^{N_B} \left[\frac{p_n^2}{2 m}
+ \frac{m \omega_n^2 q_n^2}{2} \right],
\end{eqnarray}
neglecting a contribution from the interaction term $H_I$,
which is valid for the weak interaction although
a treatment of the finite interaction is ambiguous and controversial
\cite{Hanggi08,Ingold09}.
Figures \ref{fig1}(a) and (b) show the time dependence of $u_{\eta}$
for $N_S=1$ and $10$, respectively, with $N_B=1000$ of a single DS run.
We note that although $u_{\eta}(t)$ 
rapidly oscillates, its envelope has much slower time dependence.
Periods for rapid oscillations are about 0.95 and 2.22 for
$N_S=1$ and 10, respectively: the latter value is larger than the former 
because of a larger renormalization effect due to couplings 
[the $\xi_{k \ell}$ term in Eq. (\ref{eq:A14})].
Magnitudes of time variations in $u_S(t)$ are larger than those in $u_B(t)$
because $N_S \ll N_B$.
The width of variation in $u_S(t)$ for $N_S=1$ in Fig. \ref{fig1}(a) 
is larger than that for $N_S=10$ in Fig. \ref{fig1}(b). 
Even when the energy of the system is once decreased flowing into the bath,
later it returns back to the system within the finite time \cite{Poincare}. 
Then the dissipative energy transfer 
from the system to the bath or vice versa does not occur in a long time scale
in Fig. \ref{fig1}.
This is in contrast with the result of Ref. \cite{Rosa08} which
has reported a transition from non-dissipative to dissipative 
energy transfer at $N_B \sim 300-400$ with $N_S=1$. 

In the following, we will show calculations of the stationary distributions
of the system and bath, changing $N_S$, $N_B$, interaction strength ($c_0$),
the distribution of $\omega_n$ and the ratio of $m/M$.
Hereafter the argument $u$ in the stationary distributions of $f_S(u)$ and $f_B(u)$
expresses $u=u_S$ and $u=u_B$, respectively.

\vspace{0.5cm}
\noindent
{\it 1. Effect of $N_S$}

First we study the effect of $N_S$.
Dashed, dotted, chain and solid curves in Fig. \ref{fig2}(a)
show the stationary distribution of the system $f_S(u)$ 
for $N_S=1$, 2, 5 and 10, respectively, with $N_B=100$.
$f_S(u)$ for $N_S=1$ shows an exponential-like behavior while
$f_S(u)$ for $N_S > 1$ has a structure with a peak 
near the center of the stationary distribution of the bath $f_B(u)$.
Distributions of $f_B(u)$ for $N_S=1$, 2, 5 and 10 with $N_B=100$ are plotted by 
dashed, dotted, chain and solid curves, respectively, 
in Fig. \ref{fig2}(b), which is nearly independent of $N_S$.
More detailed discussion on the $N_S$ dependence will be given in Sec. III A.

\vspace{0.5cm}
\noindent
{\it 2. Effect of $N_B$}

Calculated distributions of $f_S(u)$ for $N_S=1$ with $N_B=10$, 1000 and 1000 are 
plotted by solid, dashed and chain curves, respectively, in Fig. \ref{fig3}(a).
Similar results of $f_B(u)$ are shown in Fig. \ref{fig3}(b).
Profiles of $f_S(u)$ showing an exponential-like behavior
are almost independent of $N_B$ while those of $f_B(u)$ change: 
its width becomes narrower for larger $N_B$. 
Solid, dashed and chain curves in Fig. \ref{fig4}(a) [Fig. \ref{fig4}(b)]
show $f_S(u)$ [$f_B(u)$] for $N_B=10$, 100 and 1000, respectively,
with $N_S=10$. 
Again $f_S(u)$ of $N_S=10$ is nearly independent of $N_B$.
In particular for $N_S=N_B=10$, we obtain $f_S(u)=f_B(u)$ because
the system and bath are equivalent.
$f_B(u)$ for $N_S=10$ in Fig. 4(b) is indistinguishable to
that for $N_S=1$ in Fig. 3(b).

\vspace{0.5cm}
\noindent
{\it 3. Effect of $c_0$}

We change the coupling strength of $c_0$ in $c_{k n}=c_0/N_S N_B$.
Figure \ref{fig5}(a), (b) and (c) show distributions 
of $f_S(u)$ and $f_B(u)$ for $c_0=1.0$, 10.0 and 100.0, respectively,
with $N_S=1$ and $N_S=10$ for $N_B=100$.  
Results for $c_0=1.0$ [Fig. \ref{fig5}(a)] and $c_0=10.0$ [Fig. \ref{fig5}(b)]
are almost identical.
When $c_0$ is increased to 100.0, distribution of $f_S(u)$
becomes much wider than those in Figs. \ref{fig5}(a) and (b).
At the same time, $f_B(u)$ is modified by the stronger coupling.
We have decided to adopt $c_0=10.0$ in our DS,
related discussion being given in Sec. III A 1.

\vspace{0.5cm}
\noindent
{\it 4. Effect of distributions of $\omega_n$}

Although we have so far assumed $\omega_n=1.0$ in the bath,
we will examine additional two types of distribution ranges for $\{ \omega_n \}$:
uniform distributions in $[0.5,1.5]$ and $[2.0,3.0]$
with a fixed $\Omega_k=1.0$ in the system.
Calculated $f_S(u)$ and $f_B(u)$ for $\omega_n \in [0.5, 1.5]$ in Figs. \ref{fig6}(b) 
are almost the same as those for $\omega_n=1.0$ in Figs. \ref{fig6}(a).
In Fig. \ref{fig6}(c), where distribution of $\omega_n \in [2.0, 3.0]$ 
in the bath does not have an overlap with those of $\Omega_k=1.0$ in the system, 
$f_S(u)$ is nearly the same as those in Figs. \ref{fig6}(a) and (b)
in which the frequency ranges of the bath overlap those of the system.
In contrast, $f_B(u)$ in Fig. \ref{fig6}(c)
is quite different from those in Figs. \ref{fig6}(a) and (b) as expected.
Our results shown in Figs. \ref{fig6}(a), (b) and (c) 
suggest that $f_S(u)$ is not so sensitive to the position of frequency ranges 
of the bath relative to that of the system.  
This is in contrast with the result for $N_S=1$ in Ref. \cite{Smith08}, 
which shows that for a thermalization of the system,
the relative position between the 
oscillating frequency range of the system and that of the bath 
is very important.

\vspace{0.5cm}
\noindent
{\it 5. Effect of $m/M$}

Finally we will change a value of $m$ which has been so far assumed
to be $m=M=1.0$.
Figures \ref{fig7}(a) and (b) show $f_S(u)$ for $N_S=1$
and $N_S=10$, respectively, with $N_B=100$ for
$m/M=1.0$ (solid curves), 0.1 (dashed curves) and
0.01 (chain curves).
With decreasing the ratio of $m/M$, the magnitude at small $u$ of $f_B(u)$ is increased,
by which that of $f_S(u)$ for $N_S=1$ is slightly increased in Fig. \ref{fig7}(a).
However, the shape of $f_S(u)$ for $N_S=10$ in Fig. \ref{fig7}(b) 
is almost unchanged with changing $m/M$.

Before closing Sec. II, we point out that roles of the system 
and the bath are interchangeable in Figs. 2-7 because the Hamiltonian 
for the model A given by Eqs. (\ref{eq:A0})-(\ref{eq:A1}) has
the system$\leftrightarrow$bath symmetry for $f(t)=0$.
For example, $f_S(u)$ for $N_S=10$ and $N_B=1$ may be given by
$f_B(u)$ in Fig. 3(b) for $N_S=1$ and $N_B=10$.
Figures 2-7 show that the properties of $f_S(u)$ are mainly determined by $N_S$,
which is the main result of our study.


\section{Discussion}

\subsection{Analysis of DS results for oscillator systems}
\subsubsection{Boltzmann-Gibbs statistics}

We may theoretically evaluate the distribution of $f_S(u)$ as follows.
First we calculate the distribution for a set of variables
of $\{Q_k, V_k \}$ ($V_k=\dot{Q}_k$)
with the Boltzmann-Gibbs statistics for the infinite bath
characterized by the inverse temperature $\beta$ (see Appendix A), 
\begin{eqnarray}
f(Q,V)\:dQ dV 
&\propto& \exp\left[-\beta \sum_{k=1}^{N_S} 
\left( \frac{M V_k^2}{2} + \frac{M \Omega^2 Q_k^2}{2} \right) \right]
\Pi_{k=1}^{N_S} dQ_k dV_k,\\
&\propto& E_S^{N_S-1} e^{-\beta E_S} dE_S,
\label{eq:D1}
\end{eqnarray}
where $E_S$ denotes the energy in the system:
$E_S=\sum_{k=1}^{N_S} [MV_k^2/2+M \Omega^2 Q_k^2/2]$. From Eq. (\ref{eq:D1}),
the distribution of the system for $u$ ($=E_S/N_S$) becomes
\begin{eqnarray}
f_S(u) &=&  \frac{1}{Z} \:u^{a-1} e^{-b u} \equiv g_1(u),
\label{eq:D2}
\end{eqnarray}
with
\begin{eqnarray} 
a &=& N_S, \;\;\;\; b = N_S \beta,
\label{eq:D3} \\
Z &=& \frac{\Gamma(a)}{b^a}, 
\label{eq:D3a}
\end{eqnarray}
where $g_1(u)$ denotes the $\Gamma$ (or $\chi^2$) distribution,
its subscript 1 being attached for later purpose.
Mean ($\mu$) and variance ($\sigma^2$) of the $\Gamma$ 
distribution are given by
\begin{eqnarray}
\mu &=& \frac{a}{b}=\frac{1}{\beta}, 
\label{eq:D4} \\ 
\sigma^2 &=& \frac{a}{b^2} = \frac{1}{N_S \beta^2}.
\label{eq:D4b} 
\end{eqnarray}
Equation (\ref{eq:D4}) expresses the equipartition relation.
The distribution $f_B(u)$ of the bath for $u$ ($=E_B/N_B$)
may be obtainable in a similar way where $E_B$ signifies the bath energy.

Our DS in the preceding section has shown 
that the most influential parameter on the properties of the system is $N_S$. 
We now pay our attention to the $N_S$ dependence of
calculated means ($\mu_{\eta}$) and root-mean-square (RMS) ($\sigma_{\eta}$)
of the system ($\eta$ = S) and bath ($\eta$ = B). 
Figure \ref{fig8} shows $\mu_{\eta}$ and $\sigma_{\eta}$ 
as a function of $N_S$ with $N_B=100$ obtained by DS: 
filled (open) circles denote $\mu_S$ ($\sigma_S$) of the system: 
filled (open) squares stand for $\mu_B$ ($\sigma_B$) of the bath.
We obtain $(\mu_S, \sigma_S)=(2.84, 3.77)$, $(1.97, 2.12)$, 
$(1.302, 0.755)$ and $(1.097, 0.393)$ for $N_S=1$, 2, 5 and 10, respectively. 
With decreasing $N_S$ from $N_S=10$, $\mu_S$ and $\sigma_S$ are increased. 
In contrast, $\mu_B$ and $\sigma_B$ are almost independent of $N_S$.
An increase in $\mu_S$ with decreasing $N_S$ 
is attributed to an increase in the effective frequency 
of the system given by [Eqs. (\ref{eq:A14}) and (\ref{eq:E1})]
\begin{eqnarray}
\tilde{\Omega}^2_{kk} &=& \Omega^2_{kk}
+\frac{1}{M}\left[\frac{c_0}{N_S}
-\frac{c_0^2}{N_S^2 (N_B \omega^2+c_0/m)} \right]. \nonumber
\end{eqnarray}
An increase in $\sigma_S$ with decreasing $N_S$ is due to
an increase in $\zeta_k$ given by Eq. (\ref{eq:A16})
which is proportional to $c_0/N_S$.
When we adopt a smaller value of $c_0$, these increases
are reduced.  For example, $\mu_S$ ($\sigma_S$) calculated with a smaller $c_0=1.0$
are plotted by filled (open) triangles in Fig. 8, which shows
$(\mu_S, \sigma_S)=(1.12, 1.24)$,
(1.02, 0.78), (0.97, 0.45) and (0.97, 0.31)
for $N_S=1$, 2, 5 and 10, respectively. 
Related distributions for $N_S=1$ and 10 are plotted in Fig. 5(a).
For this choice of $c_0=1.0$,
the energy exchange between system and bath is considerably decreased. 

We have performed DS by using also
an alternative choice of couplings of $c_{kn}=c_0'/\sqrt{N_S N_B}$ 
given by Eq. (\ref{eq:E2}). 
Filled and open diamonds in Fig. 8 show
$\mu_S$ and $\sigma_S$, respectively, 
calculated with $c_0'=1/\sqrt{10}$ which is chosen such that
Eq. (\ref{eq:E2}) yields the same value of $c_{kn}=0.01$ 
as Eq. (\ref{eq:E1}) for $N_S=10$ and $N_B=100$. 
For $N_S=1$, 2, 5 and 10, we obtain
$(\mu_S, \sigma_S)=(1.61, 2.10)$, $(1.35, 1.25)$, $(1.18, 0.58)$ and
$(1.097, 0.393)$, respectively.
With decreasing $N_S$, both $\mu_S$ and $\sigma_S$ are increased, 
which are qualitatively similar to those obtained 
with couplings given by Eq. (\ref{eq:E1}).

Next we examine the profiles of $N_S$-dependent $f_S(u)$.
By using the relation between parameters $a$ and $b$ in the $\Gamma$ distribution
with its average and variance given by Eqs. (\ref{eq:D4}) and (\ref{eq:D4b}),
we may determine $a$ and $b$ by $a=(\mu^2/\sigma^2)$ and $b=\mu/\sigma^2$.
With the use of the calculated $\mu_S$ and $\sigma_S$,
we obtain $(a, b)=(0.565, 0.199)$, $(0.861, 0.437)$, $(2.98, 2.28)$ 
and $(7.78, 7.11)$ for $N_S=1$, 2, 5 and 10, respectively.
Unfortunately, these values of $a$ are not in agreement with 
the theoretical value of $a=N_S$ given by Eq. (\ref{eq:D3}). 
We have employed the $\Gamma$ distribution given by Eq. (\ref{eq:D2}) 
with the parameters $a$ and $b$ determined above
for our analysis of $f_S(u)$ shown in Fig. \ref{fig2}(a).
Dashed curves in Figs. \ref{fig9}(a)-(d) show the calculated $\Gamma$ distribution,
while solid curves express DS results.
We note in Figs. \ref{fig9}(c) and (d)
that the $\Gamma$ distributions for $N_S=5$ and 10 are ostensibly
in good agreement with calculated $f_S(u)$ although the calculated $a$
disagrees with the theoretical value of $a$ ($=N_S$) as mentioned above.
Furthermore, an agreement becomes poor for results of smaller $N_S=1$ and 2,
whose analyses will be discussed with the use of the nonextensive
statistics in the following sub-subsection.

\subsubsection{Nonextensive statistics}

\noindent
{\it a. Superstatistical approach}

A disagreement between theoretical results and DS ones 
might arise from a use of the Boltzmann-Gibbs statistics.
We will analyze the calculated results by using the nonextensive statistics
\cite{Tsallis88}-\cite{Tsallis04}.
Wilk and Wlodarczk \cite{Wilk00} and Beck \cite{Beck01} have pointed out
that the observed non-Gaussian distribution may be accounted 
for if we assume that the Gaussian distribution $e^{-\beta u}$ is averaged
over the $\Gamma$ distribution of $g(\beta)$ for fluctuating inverse 
temperature $\beta$,
\begin{eqnarray}
p(u) &\propto& e_q^{-\beta_0 u} 
= \int_0^{\infty} e^{- \beta u} \:g(\beta) \:d \beta,
\label{eq:I3}
\end{eqnarray}
with
\begin{eqnarray}
g(\beta) &=& \frac{1}{\Gamma(n/2)} 
\left(\frac{n}{2 \beta_0} \right)^{n/2} \beta^{n/2-1} 
e^{ -n\beta/2 \beta_0 }.
\label{eq:I4}
\end{eqnarray}
Here $n$ denotes the number of independent Gaussian $X_i$ contributions
to the $\chi^2$ distribution of $\beta=\sum_{i=1}^n X_i^2$ \cite{Beck01}, and
$\beta_0$ stands for the mean of $\beta$: 
$\beta_0 = \langle \beta \rangle_{\beta}$ and variance is given by 
$\langle \beta^2 \rangle_{\beta}-\beta_0^2 = (2/n) \beta_0^2$.
Equations (\ref{eq:I3}) and (\ref{eq:I4}) express the {\it superstatistics} 
whose concept may be understood such that complex nonextensive systems 
are in the nonequilibrium states 
with temporarily and spatially fluctuating inverse temperature. 

In order to more accurately account for our calculated $f_S(u)$,
we employ the concept of the superstatistics.
We assume that the $\Gamma$ distribution $f(u)$ given by Eq. (\ref{eq:D2})
is averaged over the distribution $g(\beta)$ given by Eq. (\ref{eq:I4})
with $n=2 N_S$,
\begin{eqnarray}
f_S(u) &\propto& \int_0^{\infty} u^{N_S-1} e^{- \beta N_S u} g(\beta) \:d \beta, \\
&\propto& \frac{u^{N_S-1}}{(1+ \beta_0 u)^{N_S}}. 
\end{eqnarray}
With the normalization factor, $f_S(u)$ is expressed by
the $q$-$\Gamma$ distribution $g_q(u)$,
\begin{eqnarray}
f_S(u) &=& \frac{1}{Z_q} \:u^{a-1} e_q^{-b u} \equiv g_q(u),
\label{eq:L4}
\end{eqnarray}
with
\begin{eqnarray}
q &=& 1 + \frac{1}{N_S},
\label{eq:L5b}\\
a &=& N_S, \\
b &=& N_S \beta_0,
\label{eq:L6b}
\end{eqnarray}
\renewcommand{\arraystretch}{1.5}
\begin{eqnarray}
Z_q &=& \left\{ \begin{array}{ll}
\frac{1}{[(q-1)b]^a}
\frac{\Gamma(a)\Gamma\left(\frac{1}{q-1}-a \right)}{\Gamma\left(\frac{1}{q-1} \right)}
\quad & \mbox{for $q > 1.0 $}, \\ 
\frac{\Gamma(a)}{b^a}
\quad & \mbox{for $q=1.0$},  \\
\frac{1}{[(1-q)b]^a}
\frac{\Gamma(a)\Gamma\left(\frac{1}{1-q}\right)}{\Gamma\left(\frac{1}{1-q}+a \right)}
\quad & \mbox{for $ q <1.0$}. 
\end{array} \right. 
\label{eq:L6}
\end{eqnarray}
It is easy to see that 
in the limit of $q \rightarrow 1.0$, the $q$-$\Gamma$ distribution $g_q(u)$ 
given by Eq. (\ref{eq:L4})
reduces to the $\Gamma$ distribution $g_1(u)$ given by Eq. (\ref{eq:D2}).
Average and variance of the $q$-$\Gamma$ distribution are given by
\begin{eqnarray}
\mu_q &=& \frac{a}{b[1-(q-1)(a+1)]}, 
\label{eq:L7}\\
\sigma_q^2 &=& \frac{a(2-q)}{b^2[1-(q-1)(a+2)][1-(q-1)(a+1)]^2}
\hspace{1cm}\mbox{for $q>1.0$},
\label{eq:L8}
\end{eqnarray}
which reduce to $\mu_1=a/b$ and $\sigma_1^2=a/b^2$ for $q=1.0$ in agreement 
with Eqs. (\ref{eq:D4}) and (\ref{eq:D4b}).
The $q$-$\Gamma$ distribution $g_q(u)$ has a maximum at
\begin{eqnarray}
u&=& u_m= \frac{(a-1)}{b[1-(q-1)(a-1)]}
\hspace{0.5cm}\mbox{for $a > 1.0$}.
\end{eqnarray}
The $u$ dependence of $g_q(u)$ for typical parameters
is shown in Appendix A (Fig. 15).

\vspace{0.5cm}
\noindent
{\it b. Microcanonical approach I}

Next we mention the MCA to the nonextensive statistics 
\cite{Plastino94}-\cite{Aringazin03}.
We consider microcanonical ensembles of $N$ particles with the energy $E$,
which is divided into two subsystems 1 and 2.
A probability for subsystem 1 containing $N_1$ particles to have energy $E_1$ 
is given by \cite{Plastino94,Potiguar03}
\begin{eqnarray}
f_{N_1}(E_1) &=& \frac{\Omega_1(E_1) \Omega_2(E_2)}{\Omega_{1+2}(E)},
\label{eq:X10}
\end{eqnarray}
where the structure function $\Omega_{\kappa}(E)$ ($\kappa =1, 2, 1+2$) expresses
the number of states with the energy $E$.
We assume that $\Omega_{\kappa}(E)$ is given by \cite{Plastino94,Potiguar03}, 
\begin{eqnarray}
\Omega_{\kappa}(E) &=& K m_{\kappa} \:E^{m_{\kappa}-1},
\label{eq:X11}
\end{eqnarray}
where  $K$ is a constant and $m_{\kappa}$ 
the degrees of freedom of variables in subsystem $\kappa$.
Equation (\ref{eq:X11}) is valid for ideal gases
and harmonic oscillators with $m_{\kappa} \gg 1$.

Interpreting subsystems 1 and 2 as a system and a bath, respectively,
we apply the MCA to the oscillator system
under consideration for which $m_S=N_S$ and $m_B=N_B$.
For $1 < N_S \ll N_B$ and $E_S \ll E_B$,
Eqs. (\ref{eq:X10}) and (\ref{eq:X11}) yield
\begin{eqnarray}
f_S(E_S) &\propto& E_S^{N_S-1}\left( 1- \frac{E_S}{E}\right)^{N_B-1}, \\
&=& E_S^{N_S-1}\left[1-(1-\hat{q} )\hat{\beta} E_S \right]^{1/(1-\hat{q})}, \\
&=& E_S^{N_S-1} e_{\hat{q}}^{-\hat{\beta} E_S},
\label{eq:X12}
\end{eqnarray}
with
\begin{eqnarray}
\hat{q} &=& 1- \frac{1}{(N_B-1)},
\label{eq:X13} \\
\hat{\beta} &=& \frac{N_B}{E},
\label{eq:X14}
\end{eqnarray}
where we attach hats for quantities in the MCA
to distinguish them from counterparts in the SSA. 
Equation (\ref{eq:X12}) is equivalent to the $q$-$\Gamma$ distribution
given by Eq. (\ref{eq:L4})
if we read $E_S=N_S u$ and $\hat{\beta}=\beta_0$.
Similarly, we obtain the distribution defined by \cite{Plastino94,Potiguar03}
\begin{eqnarray}
p_S(E_S) &\equiv& \frac{\Omega_B(E-E_S)}{\Omega_{S+B}(E)}, 
\label{eq:X16} \\
&\propto& e_{\hat{q}}^{- \hat{\beta} E_S}.
\label{eq:X18}
\end{eqnarray}
In the limit of $N_B \rightarrow \infty$, Eqs. (\ref{eq:X12}) and (\ref{eq:X18}) 
reduce to
\begin{eqnarray}
f_S(E_S) &\propto& E_S^{N_S-1} e^{-\hat{\beta} E_S}, \\
p_S(E_S) &\propto& e^{-\hat{\beta} E_S}, 
\label{eq:X15}
\end{eqnarray}
with
\begin{eqnarray}
\hat{\beta} &=& \frac{N_B}{E} = \frac{1}{k_B T},
\end{eqnarray}
where the equipartition relation is employed for $E_B$ ($\gg E_S$).
From Eqs. (\ref{eq:X10}) and (\ref{eq:X16}), 
a relation between $f_S(E_S)$ and $p_S(E_S)$ is given by
\begin{eqnarray}
f_S(E_S) &=&  \Omega_S(E_S) \:p_S(E_S).
\label{eq:X17}
\end{eqnarray}
With increasing $E_S$, $p_S(E_S)$ is decreased 
whereas $\Omega_S(E_S) \propto E_S^{N_S-1}$, and
then $f_S(E_S)$ has a maximum 
at $E_S=(N_S-1)/\hat{\beta}[1-(q-1)(N_S-1)]$ for $N_S > 1$. 

It should be noted that the $q$-exponential function adopted in
Refs. \cite{Almeida01}-\cite{Aringazin03} is defined by
\begin{eqnarray}
e_{q'}^{x} &=& [1+(q'-1)x]^{1/(q'-1)}
\hspace{1cm} \mbox{for $q' > 1$},
\end{eqnarray}
which is different from that given by Eq. (\ref{eq:L5}) 
proposed in Ref. \cite{Tsallis88}. The relation
between $q'$ and $q$ is $q'-1=1-q$, with which Eq. (\ref{eq:X13})
becomes $q'=1+1/(N_B-1)$ ($> 1.0$).

We have tried to apply the $q$-$\Gamma$ distribution
given by Eqs. (\ref{eq:L4})-(\ref{eq:L6}) to an analysis of profiles
of $f_S(u)$ in Fig. \ref{fig9}, but we could not obtain satisfactory results.
Rather we have phenomenologically adopted the $q$-$\Gamma$ distribution, 
choosing its parameters $a$, $b$ and $q$ such as to provide results 
in fairly good agreement with $f_S(u)$ in Fig. \ref{fig9}
with satisfying Eqs. (\ref{eq:L7}) and (\ref{eq:L8}).
Chain curves in Figs. \ref{fig9}(a) and (b) express $g_q(u)$
with $(a, b, q)=(1.0, 1.31,1.30)$ and $(1.64, 1.36, 1.09)$, respectively,
for $N_S=1$ and 2, which have been tentatively determined by a cut and try method.
It is note that $f_S(u)$ of DS is finite at $u=0.0$ for $N_S=1$, 
which requires $a=1.0$.
These chain curves are in better agreement with the calculated $f_S(u)$ 
than dashed curves expressing the $\Gamma$ distribution. 

\vspace{0.5cm}
\noindent
{c. Microcanonical approach II}

We will derive the stationary distribution with the 
alternative MCA (MCA II).
We again consider a collection of $N$ particles with the energy $E$
($=M \epsilon_0$) where $\epsilon_0$ denotes an appropriate energy unit.
A probability for its subsystem 1 containing $N_1$ particles to have energy $E_1$
($=M_1 \epsilon_0$) is given by 
\begin{eqnarray}
f_{N1}(M_1) &=& \frac{w_{N_1}(M_1)\: w_{N-N_1}(M-M_1)}{w_{N}(M)},
\label{eq:X1}
\end{eqnarray}
with
\begin{eqnarray}
w_{N}(M) &=& \frac{(M+N-1)!}{(N-1)! \: M!}.
\label{eq:X2}
\end{eqnarray}
We apply Eq. (\ref{eq:X1}) to a system plus bath
without using the condition: $1 \ll N_1 \ll N$, which is employed in the MCA I.
We assume that $M$ and $M_1$ are real as given by
\begin{eqnarray}
N &=& N_S + N_B, \;\;N_1 = N_S,  \;\;
M= \frac{E_S+E_B}{\epsilon_0}, \;\;
M_1 = \frac{E_S}{\epsilon_0}, \;\;
\label{eq:X3}
\end{eqnarray}
where $E_S$ ($E_B$) denotes an energy in the system (bath).
Then the probability for $u$ ($=E_S/N_S$) in the system is given by
\begin{eqnarray}
f_S(u) &\propto& \frac{w_{N_S}(M_1)
\: w_{N_B}(M-M_1)}{w_{N_S+N_B}(M)},
\label{eq:X4}
\end{eqnarray}
with
\begin{eqnarray}
k_B T &=& \frac{1}{\beta} = \frac{E_B}{N_B}= \mu_B,
\label{eq:X5}
\end{eqnarray}
where $\mu_B$ is the mean energy in the bath and
$w_{N}(M)$ is given by Eq. (\ref{eq:X2}) with a replacement
of $n ! \rightarrow \Gamma(n+1)$, $\Gamma(x)$ being the $\Gamma$ function.

We have calculated $f_S(u)$ with the MCA II
by using Eqs. (\ref{eq:X3})-(\ref{eq:X5}), whose results
with $\epsilon_0=1.0$ (dashed curves), 0.1 (dotted curves) 
and 0.01 (chain curves) are shown in Figs. \ref{fig10}(a)-(d).
With decreasing $\epsilon_0$, results of MCA II are expected to
approach the classical limit.
Although a general trend is accounted for by MCA II calculations,
their agreement with DS results is not so good.

\subsection{Comparison with ideal-gas systems}

Our model A given by Eqs. (\ref{eq:A0})-(\ref{eq:A1}) may be applied 
to ideal gases (the system) coupled to finite bath, for which we set $V(Q_k)=0$. 
We have performed DS with the same model parameters
(except for $\Omega_k=0$) as in the case of oscillator systems mentioned in Sec. II.
Solid curves in Figs. \ref{fig11}(a), (b), (c) and (d) show calculated $f_S(u)$ of 
$N_S$-body ideal gases for $N_S=1$, 2, 5 and 10, respectively, with $N_B=100$.
For a comparison, we show by dotted curves, the corresponding results
for oscillators having been plotted in Fig. \ref{fig2}(a). 
The energy distributions of bath, $f_B(u)$, for ideal-gas
systems are almost the same 
as those for oscillator systems shown in Fig. \ref{fig2}(b).
Comparing solid curves to dotted curves, we note that the distribution
of $f_S(u)$ for ideal gases has larger magnitude at small $u$
than that for oscillators. This yields the smaller average energy
in ideal gases than that in oscillators, which is related with the fact
the former has a smaller degree of freedom than the latter, 
as expressed in the equipartition relation.

Figure 12 shows the $N_S$ dependence of $\mu_{\eta}$
and $\sigma_{\eta}=S, B$:
filled (open) circles show $\mu_S$ ($\sigma_S$) and
filled (open) squares denote $\mu_B$ ($\sigma_B$). 
Although $\mu_S$ in Fig. 12 has similar $N_S$ dependence to 
that in Fig. 8 for oscillator systems, 
magnitudes of the former are smaller than the latter. 
The ratio of $\mu_S (IG)/\mu_S(OSC)$ approaches 0.5 with increasing $N_S$,
although the ratio is increased for $N_S \rightarrow 1$.

We have analyzed calculated $f_S(u)$ in Figs. \ref{fig11}(a)-(d) 
by using the $\Gamma$ distribution given by Eq. (\ref{eq:D2}) with
\begin{eqnarray}
a &=& \frac{N_S}{2}, \;\;\;\; b=N_S \beta,
\label{eq:D6}
\end{eqnarray}
which lead to
\begin{eqnarray}
\mu&=& \frac{a}{b}=\frac{1}{2 \beta}, 
\label{eq:D7} \\ 
\sigma^2 &=& \frac{a}{b^2}=\frac{1}{2 N_S \beta^2}.
\label{eq:D7b}
\end{eqnarray}
Equation (\ref{eq:D7}) expresses the equipartition relation of ideal gases.
Simulations shown in Fig. 11 yield $(\mu_S, \sigma_S)=$ (2.59, 3.62), (1.50, 1.76),
(0.716, 0.504) and (0.469, 0.222) for $N_S=1$, 2, 5 and 10, respectively,
from which we obtain $(a, b)=$ (0.514 0.198), (0.733, 0.487), 
(2.02, 2.82) and (4.47 9.54).
Dashed curves in Figs. \ref{fig11}(a)-(d) express the $\Gamma$ distribution
calculated with the use of $a$ and $b$ thus obtained.
They are in good agreement with the DS results
for $N_S=5$ and 10, but not for $N_S=1$ and 2.  
Chain curves express the $q$-$\Gamma$ distributions
obtained with $(a, b, q)=$ $(0.61, 0.38, 1.18)$ and $(1.0, 0.85, 1.09)$ for
$N_S=1$ and 2, respectively, which have been determined by a cut-and-try method.
Results of the $q$-$\Gamma$ distribution are in better agreement
with DS than those of the $\Gamma$ distribution. This situation is
the same as in the case of oscillator systems as discussed in Sec. III A.

\subsection{Bath containing coupled oscillators}

In most of existing models for open systems
\cite{Caldeira81}-\cite{Rosa08},
baths are assumed to be consisting of uncoupled oscillators. 
In order to study the effect of couplings of oscillators in a bath, 
we consider the models B and C in which baths
consist of coupled oscillators with the periodic and fixed-end
boundary conditions, respectively. 

\vspace{0.5cm}
\noindent
{\it 1. Model B}

In the model B, we assume that the Hamiltonian is given 
by Eqs. (\ref{eq:A0})-(\ref{eq:A4}) with $v(q_n)$,
\begin{eqnarray}
v(q_n) &=& \frac{b}{2} (q_n-q_{n+1})^2
\hspace{1cm}\mbox{(model B)},
\label{eq:B1}
\end{eqnarray}
under the periodic boundary condition:
\begin{eqnarray}
q_{N_B+n}&=& q_{n}, \;\;p_{N_B+n}= p_{n},
\label{eq:B5}
\end{eqnarray}
where $b$ denotes the spring constant between neighboring
sites in the bath and $N_B$ is assumed even without a loss of generality.

Equations of motion for $Q_k$ and $q_n$ are given by
\begin{eqnarray}
M \ddot{Q}_k &=& - V'(Q_k) 
- \sum_{n=-N_B/2}^{N_B/2-1} c(Q_k-q_n)+f(t), 
\label{eq:B6}\\
m \ddot{q}_n &=& -b(2 q_n-q_{n-1}-q_{n+1})- \sum_{k=1}^{N_S} c(q_n-Q_k).
\label{eq:B7}
\end{eqnarray}

By using a transformation mentioned in Appendix B, we obtain
the Langevin equation for $Q_k(t)$ given by Eq. (\ref{eq:A13}) with
\begin{eqnarray}
M \xi_{k \ell} &=& c N_B \delta_{k \ell}
- \frac{c^2 N_B}{m \tilde{\omega}_0^2},
\label{eq:B16}\\
\gamma_{k \ell}(t) &=& \left( \frac{c^2 N_B}{m \tilde{\omega}_0^2} \right)
\cos \tilde{\omega}_0 t,
\label{eq:B22}\\
\zeta_k(t) &=& c \sqrt{N_B}
\left[ \tilde{q}_0(0) \cos \tilde{\omega}_0 t 
+ \frac{\dot{\tilde{q}}_0(0)}{\tilde{\omega}_0}\sin \tilde{\omega}_0 t
\right],
\label{eq:B17}
\end{eqnarray}
with
\begin{eqnarray}
\tilde{\omega}_0^2 &=& \frac{c N_S}{m}.
\label{eq:B18}
\end{eqnarray}

\vspace{0.5cm}
\noindent
{2. \it Model C}

In the model C, we assume that the Hamiltonian is given 
by Eqs. (\ref{eq:A0})-(\ref{eq:A4}) with $v(q_n)$,
\begin{eqnarray}
v(q_n)&=& \frac{b}{2} (q_n-q_{n+1})^2
\hspace{1cm}\mbox{(model C)},
\label{eq:C1}
\end{eqnarray}
under the fixed-end boundary condition given by
\begin{eqnarray}
q_{0}&=& q_{N_B+1}=0, \;\;p_{0}= p_{N_B+1}=0.
\label{eq:C5}
\end{eqnarray}

Equations of motion for $Q_k$ and $q_n$ are given by
\begin{eqnarray}
M \ddot{Q}_k &=& - V'(Q_k) - \sum_{n=0}^{N_B+1} c(Q_k-q_n)+f(t), 
\label{eq:C6}\\
m \ddot{q}_n &=& -b(2 q_n-q_{n-1}-q_{n+1})- \sum_{k=1}^{N_S} c(q_n-Q_k).
\label{eq:C7}
\end{eqnarray}

By using a transformation mentioned in Appendix C, we obtain
the Langevin equation given by Eq. (\ref{eq:A13}) with
\begin{eqnarray}
M \xi_{k \ell} &=& c(N_B+2) \delta_{k \ell}
- \sum_{s=1}^{N_B} \frac{c^2 a_s^2}{m \hat{\omega}_s^2}, 
\label{eq:C17}\\
\gamma_{k \ell}(t) &=& \sum_{s=1}^{N_B}
\left( \frac{c^2 a_s^2}{m \hat{\omega}_s^2} \right)
\:\cos \hat{\omega}_s t, 
\label{eq:C18}\\ 
\zeta_k(t)&=& \sum_{s=1}^{N_B} 
\:c \:a_s \left[ \hat{q}_s(0) \cos \hat{\omega}_s t
+ \frac{\dot{\hat{q}}_s(0)}{\hat{\omega}_s} \sin \hat{\omega}_s t
\right],
\label{eq:C19}
\end{eqnarray}
where $\hat{\omega}_s$ and $a_s$ are expressed by
\begin{eqnarray}
\hat{\omega}_s^2 &=& \omega_s^2+ \frac{c (N_S+2)}{m}, \\
a_s &=& \sqrt{\frac{1}{2(N_B+1)}}
\left[
\cos \left(\frac{\pi s}{2}-\frac{(N_B+2) \pi s}{2(N_B+1)} \right)  
-\cos \left(\frac{\pi s}{2}+\frac{(N_B+2) \pi s}{2(N_B+1)} \right)
\right] \nonumber \\
&& \times {\rm cosec} \left( \frac{\pi s}{2(N_B+1)} \right).
\label{eq:C20}
\end{eqnarray}

DS calculations for models B and C have been performed for oscillator systems 
with the same parameters as in Sec. II in addition to $b=1.0$.
Dashed and solid curves in Fig. \ref{fig13}(a) [Fig. \ref{fig13}(b)]
show $f_S(u)$ [$f_B(u)$] of the model B
for $N_S=1$ and 10, respectively, with $N_B=100$.
Dashed and solid curves in Fig. \ref{fig14}(a) [Fig. \ref{fig14}(b)]
show $f_S(u)$ [$f_B(u)$] of the model C
for $N_S=1$ and 10, respectively, with $N_B=100$.
Profiles of $f_S(u)$ and $f_B(u)$ of the model B in Fig. \ref{fig13} 
are similar to those of the model C in Fig. \ref{fig14}.
Comparing Figs. \ref{fig13} and \ref{fig14} with Fig. \ref{fig2},
we note that couplings in oscillators of the bath have essentially no
effects on the behavior of $f_S(u)$ of the system, although they have
some effects on $f_B(u)$ as expected.

\subsection{Comparisons among various models}

Table 1 summarizes comparisons among elements of $\xi_{kn}$, $\gamma_{kn}$
and $\zeta_k$ in Langevin equations
derived from various models for open systems 
including CL \cite{Caldeira81} and MK models \cite{Ford87} and
models A, B and C which are proposed in Secs. II and III.
Additional interactions $\xi_{kn}$ induced by introduced couplings
between the system and bath remain finite in the models A, B and C
although they vanish in the CL and MK models for $N_S=1$.
We note that functional forms of $\xi_{k\ell}$ and $\zeta_k$
in all the models are similar.
This is the reason why properties of $f_S(u)$ and $f_B(u)$
in Figs. \ref{fig2}, \ref{fig13} and $\ref{fig14}$ are similar.
We note, however, that the kernel $\xi_{k\ell}$ of the model B
is oscillating and not dissipative even for $N_B \rightarrow \infty$,
which arises from the translational symmetry in the bath.

\begin{table}
\begin{center}
\caption{Terms of $\xi_{k\ell}$, $\gamma_{k \ell}$ and $\zeta_k$ 
in the Langevin equation,
$M \ddot{Q}_k(t) = -V'(Q_k)- M \sum_{\ell} \xi_{k\ell} Q_{\ell}(t)
- \sum_{\ell} \int_0^t \gamma_{k\ell}(t-t') \dot{Q}_{\ell}(t')\:dt'
- \sum_{\ell} \gamma_{k\ell}(t) Q_{\ell}(0) + \zeta_k(t)$,
calculated by various models: 
1) CL model [Eq. (\ref{eq:I1})]: 2) MK model [Eq. (\ref{eq:J1})]: 
3) the model A [Eq. (\ref{eq:A1})]: 4) the model B [Eq. (\ref{eq:B1})]: 
5) the model C [Eq. (\ref{eq:C1})].
The CL and MK models are for $N_S=1$ for which subscripts
$k, \ell$ are dropped. 
}
\renewcommand{\arraystretch}{1.5}
\begin{tabular}{|c||c|c|c|} \hline
model
& M $\xi_{k \ell}$ & $\gamma_{k \ell}$ 
& $\zeta_k$ 
\\ \hline \hline
CL$^{1)}$ ($N_S=1$) &  0
& $\sum_{n} \left( \frac{c_n^2}{m \omega_n^2}\right) \cos \omega_n t$
& $\sum_{n}  c_n 
\left[q_n(0) \cos \omega_n t
+\left(\frac{\dot{q}_n(0)}{\omega_n}\right) \sin \omega_n t
\right]$ 
\\ \hline
MK$^{2)}$ ($N_S=1$) & 0
& $\sum_{n} m \omega_n^2 \cos \omega_n t$
& $\sum_{n} m \omega_n^2  
\left[q_n(0) \cos \omega_n t
+\left(\frac{\dot{q}_n(0)}{\omega_n}\right) \sin \omega_n t
\right]$ 
\\ \hline
A$^{3)}$  &  $\sum_n \left[ c_{kn}\delta_{k \ell} 
-\frac{c_{kn} c_{\ell n}}{m \tilde{\omega}_n^2}
\right]$
& $ \sum_{n} \left( \frac{c_{kn} c_{\ell n}}{m \tilde{\omega}_n^2} \right)
\cos \tilde{\omega} t$
& $\sum_{n} c_{k n} 
\left[q_n(0) \cos \tilde{\omega}_n t
+\left(\frac{\dot{q}_n(0)}{\tilde{\omega}_n}\right) \sin \tilde{\omega}_n t
\right]$ 
\\ \hline
B$^{4)}$  &  $c N_B \delta_{k \ell}
- \frac{c^2 N_B}{m \tilde{\omega}_0^2}$
& $\left( \frac{c^2 N_B}{m \tilde{\omega}_0^2}\right) \cos \tilde{\omega}_0^2 t$
& $ c \sqrt{N_B} 
\left[\tilde{q}_0(0) \cos \tilde{\omega}_0 t
+\left(\frac{\dot{\tilde{q}}_0(0)}{\tilde{\omega}_n}\right) 
\sin \tilde{\omega}_n t
\right]$  
\\ \hline
C$^{5)}$  &  $c(N_B+2)\delta_{k\ell}
- \sum_{s} \frac{c^2 a_s^2}{m \hat{\omega}_s^2}$
& $\sum_{s} \left( \frac{c^2 a_s^2}{m \hat{\omega}_s^2}\right) 
\cos \hat{\omega}_s t$
& $\sum_{s} c a_s 
\left[\hat{q}_s(0) \cos \hat{\omega}_s t
+\left(\frac{\dot{\hat{q}}_s(0)}{\hat{\omega}_s}\right) \sin \hat{\omega}_s t
\right]$ 
\\ \hline

\end{tabular}
\end{center}
\end{table}

\section{Concluding remarks}

It is worthwhile to make a comparison between the SSA and MCA,
which lead to equivalent $q$-$\Gamma$ distributions 
given by Eqs. (\ref{eq:L4}) and (\ref{eq:X12}).
We should, however, note that
the entropic index of $f_S(u)$ obtained in the SSA [Eq. (\ref{eq:L5b})] 
is different from that derived in the MCA 
[Eq. (\ref{eq:X13})] as shown by Eq. (\ref{eq:L10}): 
$q$ in the SSA is expressed in terms of $N_S$ and greater than unity, 
while $q$ in the MCA is expressed in terms of $N_B$ and less than unity. 
Our DS has  shown that $f_S(u)$ depends on $N_S$ in Fig. \ref{fig2} or \ref{fig9} 
while it is almost independent of $N_B$ in Figs. \ref{fig3} and \ref{fig4}, 
which suggests that the entropic index of $f_S(u)$ depends mainly on $N_S$ 
but only weakly on $N_B$. Furthermore, our phenomenological analyses show 
that the deduced entropic indexes are greater than unity.
These facts seem to support the SSA \cite{Wilk00,Beck01} 
but throw doubt on the MCA and its applications \cite{Plastino94}-\cite{Aringazin03}, 
although more detailed study is necessary to draw a definite conclusion.

To summarize, we have studied the properties of classical small systems coupled to
finite bath, by employing the ($N_S+N_B$) models A, B and C, 
in which $N_S$-body system is coupled to $N_B$-body bath.
Simulations for oscillator and ideal-gas systems have shown the following:

\noindent
(i) the energy of the system oscillates rapidly 
although its envelope has much slower time dependence,

\noindent
(ii) the dissipation of the system energy is not observed
in our DS with $N_S \sim 1-10$ and $N_B \sim 10-1000$,

\noindent
(iii) the stationary energy distribution of the system $f_S(u)$
for $N_S > 1$ has a peak at about the average energy of the bath,
although $f_S(u)$ for $N_S=1$ has an exponential-like distribution
decreasing monotonously with increasing $u$,

\noindent
(iv) calculated $f_S(u)$, whose properties depend mainly on $N_S$
but only weakly on $N_B$, may be phenomenologically described
by the $\Gamma$ or $q$-$\Gamma$ distribution [Eq. (\ref{eq:L4})], and

\noindent
(v) the coupling among oscillators in the bath yields little
effect in classical systems.

\noindent
The item (i) is consistent with a previous study for $N_S=1$ 
in Ref. \cite{Rosa08}. 
The item (ii) suggests that for the energy dissipation of system,
we might need to adopt a much larger $N_B$ ($\gg 1000$) \cite{Poincare}. 
The thermalized state reported 
in Refs. \cite{Smith08,Wei09} corresponds to our state
for $N_S=1$ with the exponential-like distribution, in agreement with the item (iii).
The item (iv) is favorable to the SSA but not to the MCA although 
either of them cannot {\it quantitatively} explain the DS results.
The item (v) is consistent with the classical specific heat of harmonic oscillators
for which both Einstein and Debye models yield the same results. 
Our model A given by Eqs. (\ref{eq:A0})-(\ref{eq:A1})
is expected to have a wide applicability to classical small systems: 
for example, for studies on a system with various potentials
$V(Q)$ like the bi-stable potentials and 
on a work performed by time-dependent external force $f(t)$ in Eq. (\ref{eq:A2}).
These subjects are left as our future study.

\begin{acknowledgments}
This work is partly supported by
a Grant-in-Aid for Scientific Research from 
Ministry of Education, Culture, Sports, Science and Technology of Japan.  
\end{acknowledgments}

\appendix*

\section{A. $q$-$\chi^2$ and $q$-$\Gamma$ distributions}
\renewcommand{\theequation}{A\arabic{equation}}
\setcounter{equation}{0}

\subsection{The $q$-$\chi^2$ distribution}
We will show that if $n$ independent variables of $\{x_i \}$ follow 
the $q$-Gaussian distribution, a variable defined by $Y=\sum_{i=1}^n x_i^2$
follows the $q$-$\chi^2$ distribution with rank $n$ defined by
\begin{eqnarray}
P(Y) &= & \frac{1}{Z} \:e_q^{-Y} Y^{n/2-1},
\label{eq:Z1}
\end{eqnarray}
where $Z$ stands for the normalization factor.

In order to derive Eq. (\ref{eq:Z1}), we first define a new
variable of $X^2=\sum_{i=1}^n x_i^2$, for which we obtain
\begin{eqnarray}
p(x) dx &\propto & e_q^{-\sum_i x_i^2} \prod_{i=1}^n dx_i, \nonumber \\
&\propto& e_q^{-X^2} X^{n-1} dX, \nonumber \\
&\propto& e_q^{-Y} Y^{(n-1)/2} Y^{-1/2} dY, \nonumber \\
&=& e_q^{-Y} Y^{n/2-1} dY,
\label{eq:Z2}
\end{eqnarray}
leading to the $q$-deformed $\chi^2$ distribution given by Eq. (\ref{eq:Z1}).

It is noted that the factorization is not satisfied for 
the $q$-exponential function \cite{Tsallis88,Hasegawa10},
\begin{eqnarray}
e_q^{- \sum_i x_1^2} &\neq& \prod_i e_q^{-x_i^2},
\end{eqnarray}
except for $q=1$ or $n=1$. Then we cannot employ the 
method of the characteristic function by which
the $\chi^2$-function is conventionally derived from $n$ independent
Gaussian. 

\subsection{The $q$-$\Gamma$ distribution}
When generalizing $n/2$ in Eq. (\ref{eq:Z1}) to a real number $a$, we obtain 
the $q$-$\Gamma$ distribution,
\begin{eqnarray}
g_q(u) &=& \frac{1}{Z_q} \:u^{a-1} e_q^{-b u},
\label{eq:Z3}
\end{eqnarray}
where $Z_q$ is given by Eq. (\ref{eq:L6}).
Some numerical examples of $g_q(u)$ are shown in Fig. \ref{fig15}. 
The $q$-$\Gamma$ distribution for $q > 1.0$
has a larger magnitude than the $\Gamma$ distribution ($q=1.0$) at large $u$
because of the flat-tail properties of the $q$-exponential function \cite{Tsallis88}. 
In contrast, $q$-$\Gamma$ distribution for $q < 1.0$
has a compact structure because of cut-off properties of the $q$-exponential function
with no magnitudes for $u \geq 1/(1-q) b$.

\section{B. Langevin equation in the model B}
\renewcommand{\theequation}{B\arabic{equation}}
\setcounter{equation}{0}
We will explain a derivation of the Langevin equation in the
model B given by Eqs. (\ref{eq:A0})-(\ref{eq:A4}), (\ref{eq:B1}) and (\ref{eq:B5}).
By using the transformation given by \cite{Plyukhin01,Florencio85}
\begin{eqnarray}
q_n &=& \frac{1}{\sqrt{N_B}} \sum_{s=-N_B/2}^{N_B/2-1}
\:e^{i(2 \pi n s/N_B)} \tilde{q}_s, 
\label{eq:B8}\\
p_n &=&  \frac{1}{\sqrt{N_B}}
\sum_{s=-N_B/2}^{N_B/2-1} \:e^{i(2 \pi n s/N_B)} \tilde{p}_s,
\label{eq:B9}
\end{eqnarray}
we obtain the diagonalized $H_B$,
\begin{eqnarray}
H_B &=& \sum_{s=-N_B/2}^{N_B/2-1}
\left( \frac{1}{2m} \tilde{p}_s^* \tilde{p}_s
+ \frac{m \omega_s^2}{2} \tilde{q}_s^* \tilde{q}_s \right),
\label{eq:B10}
\end{eqnarray}
with
\begin{eqnarray}
\omega_s^2 &=& \left( \frac{4 b}{m}\right) \sin^2 \left(\frac{\pi s}{N_B}  \right)
\hspace{1cm}\mbox{$(s=-N_B/2, \cdot\cdot\cdot, N_B/2-1)$}.
\label{eq:B11}
\end{eqnarray}
Substituting Eqs. (\ref{eq:B8}) and (\ref{eq:B9}) to Eq. (\ref{eq:A4})
lead to
\begin{eqnarray}
H_{I} &=& \frac{c N_B}{2} \sum_{k=1}^{N_S} Q_k^2
+ \frac{c N_S}{2}\sum_{s=-N_B/2}^{N_B/2-1}
\tilde{q}_s^*\tilde{q}_s 
- c \sqrt{N_B} \tilde{q}_0 \sum_{k=1}^{N_S} Q_k.
\label{eq:B12}
\end{eqnarray}

Then equations of motion become
\begin{eqnarray}
M \ddot{Q}_k &=& -V'(Q_k)-c N_B Q_k
+ c \sqrt{N_B} \tilde{q}_0 + F(t), 
\label{eq:B13} \\
m \ddot{\tilde{q}}_s &=& -m \tilde{\omega}_s^2 \tilde{q}_s
+ c \sqrt{N_B} \:\sum_{k-1}^{N_S} Q_k \delta_{s 0},
\label{eq:B14}
\end{eqnarray}
with
\begin{eqnarray}
\tilde{\omega}_s^2 &=& \omega_s^2+\frac{c N_S}{m}.
\label{eq:B15}
\end{eqnarray}
Note that the third term of Eq. (\ref{eq:B13}) 
and the second term of Eq. (\ref{eq:B14})
include only the $s=0$ component.
Substituting a formal solution of $\tilde{q}_s$ to Eq. (\ref{eq:B13}),
we obtain the Langevin equation given by Eqs. (\ref{eq:A13}) 
and (\ref{eq:B16})-(\ref{eq:B18}).

\section{C. Langevin equation in the model C}
\renewcommand{\theequation}{C\arabic{equation}}
\setcounter{equation}{0}

A derivation of the Langevin equation in the
model C given by Eqs. (\ref{eq:A0})-(\ref{eq:A4}),(\ref{eq:C1}) and (\ref{eq:C5})
will be explained.
A transformation given by \cite{Plyukhin01,Florencio85}
\begin{eqnarray}
q_n &=& \sqrt{\frac{2}{N_B+1}} \sum_{s=1}^{N_B}
\:\sin\left( \frac{\pi n s}{N_B+1} \right) \hat{q}_s,
\label{eq:C8}\\
p_n &=& \sqrt{\frac{2}{N_B+1}} \sum_{s=1}^{N_B}
\:\sin\left( \frac{\pi n s}{N_B+1} \right) \hat{p}_s,
\label{eq:C9}
\end{eqnarray}
yields the diagonalized $H_B$,
\begin{eqnarray}
H_B &=& \sum_{s=1}^N \left( \frac{\hat{p}_s^2}{2m}
+ \frac{m \omega_s^2 \hat{q}_s^2}{2} \right),
\label{eq:C10}
\end{eqnarray}
with
\begin{eqnarray}
\omega_s^2 &=& \left( \frac{4b}{m} \right) 
\sin^2 \left[ \frac{\pi s}{2(N_B+1)} \right]
\hspace{1cm}\mbox{$(s=1,2,\cdot \cdot N_B)$}
\label{eq:C11}
\end{eqnarray}
From a transformation given by Eqs. (\ref{eq:C8}) and (\ref{eq:C9}), 
we obtain $H_{I}$ given by
\begin{eqnarray}
H_{I} &=& \frac{(N_B+2)c}{2} \sum_{k=1}^{N_S} Q_k^2
+ \frac{(N_S+2)c}{2} \sum_{s=1}^{N_B} \hat{q}_s^2
-c \sum_{k=1}^{N_S} Q_k \sum_{s=1}^{N_B} \:a_s \hat{q}_s,
\label{eq:C12} 
\end{eqnarray}
with
\begin{eqnarray}
a_s &=& \sqrt{\frac{2}{N_B+1}} 
\sum_{n=0}^{N_B+1} \sin\left( \frac{\pi n s}{N_B+1} \right).
\label{eq:C13}
\end{eqnarray}

Then equations of motion for $Q_k$ and $\hat{q}_s$ become
\begin{eqnarray}
M \ddot{Q}_k &=& - V'(Q_k) -c(N_B+2)Q_k
+  c \sum_{s=1}^{N_B} \:a_s \hat{q}_s+ F(t), 
\label{eq:C14}\\
m \ddot{\hat{q}}_s &=& - m \hat{\omega}_s^2  \hat{q}_s 
+ c a_s \sum_{k=1}^{N_S}\:Q_k,
\label{eq:C15}
\end{eqnarray}
with
\begin{eqnarray}
m \hat{\omega}_s^2 &=& m \omega_s^2+ c (N_S+2),
\label{eq:C16}
\end{eqnarray}
Substituting a formal solution of $\hat{q}_s$ to Eq. (\ref{eq:C14}),
we obtain the Langevin equation given by Eqs. (\ref{eq:A13}) 
and (\ref{eq:C17})-(\ref{eq:C19}).

The $s$ dependence of $a_s$ given by Eq. (\ref{eq:C13}) or Eq. (\ref{eq:C20})
is plotted in Fig. \ref{fig16},
showing the zig-toothed structure whose magnitude decreases
rapidly with increasing $s$. 


\newpage
\begin{figure}
\begin{center}
\end{center}
\caption{
(Color online) Time dependences of $u_{S}(t)$ and $u_B(t)$
for (a) $N_S=1$ and (b) $N_S=10$ with $N_B=1000$ (a single DS run),
inset showing enlarged plots of $u_{S}(t)$ for $t=0$ to 60.
}
\label{fig1}
\end{figure}

\begin{figure}
\begin{center}
\end{center}
\caption{
(Color online) 
(a) Stationary distributions of (a) $f_S(u)$ and (b) $f_B(u)$ with $N_B=100$ 
for various $N_S$: $N_S=1$ (dashed curves), 2 (dotted curves), 5 (chain curves)
and 10 (solid curves). 
}
\label{fig2}
\end{figure}

\begin{figure}
\begin{center}
\end{center}
\caption{
(Color online) Stationary distributions of (a) $f_S(u)$ and (b) $f_B(u)$ 
with $N_S=1$ for various $N_B$: $N_B=10$ (solid curves), 100 (dashed curves) 
and 1000 (chain curves). 
}
\label{fig3}
\end{figure}

\begin{figure}
\begin{center}
\end{center}
\caption{
(Color online) Stationary distributions of (a) $f_S(u)$ and (b) $f_B(u)$ 
with $N_S=10$ for various $N_B$: $N_B=10$ (solid curves), 100 (dashed curves) 
and 1000 (chain curves). 
}
\label{fig4}
\end{figure}

\begin{figure}
\begin{center}
\end{center}
\caption{
(Color online) 
Stationary distributions of $f_S(u)$ and $f_B(u)$ 
for (a) $c_0=1.0$, (b) 10.0 and (c) 100.0 in $c_{kn}=c_o/N_S N_B$:
chain (solid) curve denotes $f_S$ for $N_S=1$ ($N_S=10$),
and dotted (dashed) curve expresses $f_B$ for $N_S=1$ ($N_S=10$) with $N_B=100$,
$f_B(u)$ being divided by a factor of two. 
}
\label{fig5}
\end{figure}

\begin{figure}
\begin{center}
\end{center}
\caption{
(Color online) Stationary distributions of
$f_S(u)$ (solid curves) and $f_B(u)$ (dashed curves)
for (a) $\omega_n=1.0$, (b) $\omega_n \in [0.5,1.5]$ and (c) $\omega_n \in [2.0,3.0]$
with $N_S=10$ and $N_B=100$, $f_B(u)$ being divided by a factor of two. 
}
\label{fig6}
\end{figure}

\begin{figure}
\begin{center}
\end{center}
\caption{
(Color online) Stationary distributions of
$f_S(u)$ for (a) $N_S=1$ and (b) $N_S=10$ with $N_B=100$
for various $m/M$: $m/M=1.0$ (solid curves), $0.1$ (dashed curves)
and 0.01 (chain curves).
}
\label{fig7}
\end{figure}

\begin{figure}
\begin{center}
\end{center}
\caption{
(Color online) 
$N_S$ dependences of $\mu_{\eta}$ and $\sigma_{\eta}$ 
of systems ($\eta$= S) and baths ($\eta$= B) with $N_B=100$:
filled (open) circles show $\mu_S$ ($\sigma_S$), 
and filled (open) squares $\mu_B$ ($\sigma_B$) with $c_0=10.0$: 
filled (open) triangles express $\mu_S$ ($\sigma_S$) calculated with $c_0=1.0$:
filled (open) diamonds denote $\mu_S$ ($\sigma_S$) calculated 
with the coupling given by $c_{kn}=1.0/\sqrt{10 N_S N_B}$ [Eq. (\ref{eq:E2})] (see text).
}
\label{fig8}
\end{figure}

\begin{figure}
\begin{center}
\end{center}
\caption{
(Color online) 
The $u$ dependence of $f_S(u)$ for (a) $N_S=1$,
(b) $N_S=2$, (c) $N_S=5$ and (d) $N_S=10$ with $N_B=100$ obtained by 
our direct simulation (DS: solid curves) and the $\Gamma$ distribution ($\Gamma$)
given by Eq. (\ref{eq:D2}) (dashed curved). Chain curves in (a) and (b)
express the $q$-$\Gamma$ distribution ($q$-$\Gamma$) given by Eq. (\ref{eq:L4}) (see text).
}
\label{fig9}
\end{figure}

\begin{figure}
\begin{center}
\end{center}
\caption{
(Color online) 
The $u$ dependence of $f_S(u)$ for (a) $N_S=1$,
(b) $N_S=2$, (c) $N_S=5$ and (d) $N_S=10$ with $N_B=100$ obtained 
by the MCA II with $\epsilon_0=1.0$ (dashed curves), 0.1 (dotted curves) 
and 0.01 (chain curves) [Eqs. (\ref{eq:X3})-(\ref{eq:X5})],
solid curves expressing DS results.
}
\label{fig10}
\end{figure}

\begin{figure}
\begin{center}
\end{center}
\caption{
(Color online) 
The $u$ dependence of $f_S(u)$ of ideal-gas systems
for (a) $N_S=1$, (b) $N_S=2$, (c) $N_S=5$ and (d) $N_S=10$ with $N_B=100$:
DS (IG: solid curves), the $\Gamma$ distribution ($\Gamma$ (IG): dashed curves)
and the $q$-$\Gamma$ distribution ($q$-$\Gamma$ (IG): chain curves).
For a comparison, DS results for oscillator system (OSC)
are plotted by dotted curves (see text). 
}
\label{fig11}
\end{figure}

\begin{figure}
\begin{center}
\end{center}
\caption{
(Color online) 
$N_S$ dependences of $\mu_{\eta}$ and $\sigma_{\eta}$
of ideal gas systems ($\eta$= S: circles) and 
baths ($\eta$= B: squares) with $N_B=100$: 
filled and open marks denote mean and RMS, respectively.
}
\label{fig12}
\end{figure}

\begin{figure}
\begin{center}
\end{center}
\caption{
(Color online) Stationary distributions of
(a) $f_S(u)$ and (b) $f_B(u)$ of the model B 
for $N_S=1$ (dashed curves) and 10 (solid curves) with $N_B=100$.
}
\label{fig13}
\end{figure}

\begin{figure}
\begin{center}
\end{center}
\caption{
(Color online) Stationary distributions of
(a) $f_S(u)$ and (b) $f_B(u)$ of the model C 
for $N_S=1$ (dashed curves) and 10 (solid curves) with $N_B=100$. 
}
\label{fig14}
\end{figure}

\begin{figure}
\begin{center}
\end{center}
\caption{
(Color online) 
The $q$-$\Gamma$ distribution $g_q(u)$ [Eq. (\ref{eq:Z3})]
for (a) $a=0.5$, (b) 1.0, (c) 1.5, (d) 2.0, (e) 3.0 and (f) 4.0 with $b=1.0$:
$q=0.9$ (chain curves), 1.0 (dashed curves) and 1.1 (solid curves). 
}
\label{fig15}
\end{figure}

\begin{figure}
\begin{center}
\end{center}
\caption{
(Color online) The $s$ dependence of $a_s$ for $N=10$ (dashed curve), 20 (chain curve), 
50 (dotted curve) and 100 (solid curve) 
[Eq. (\ref{eq:C13}) or (\ref{eq:C20})].
}
\label{fig16}
\end{figure}

\end{document}